\documentclass[apj, iop]{emulateapj}
\usepackage{natbib}
\usepackage{rotating}
\def\simlt{\mathrel{\rlap{\lower 3pt\hbox{$\sim$}}\raise 2.0pt\hbox{$<$}}}
\def\simgt{\mathrel{\rlap{\lower 3pt\hbox{$\sim$}} \raise 2.0pt\hbox{$>$}}}
\def\spose#1{\hbox to 0pt{#1\hss}}
\newcommand{\lta}{\mathrel{\spose{\lower 3pt\hbox{$\mathchar"218$}}
      \raise 2.0pt\hbox{$\mathchar"13C$}}}
\newcommand{\gta}{\mathrel{\spose{\lower 3pt\hbox{$\mathchar"218$}}
      \raise 2.0pt\hbox{$\mathchar"13E$}}}

\newcommand{\beq}{\begin{equation}}
\newcommand{\eeq}{\end{equation}}

\begin{document}

\title{Observability of dual active galactic nuclei in merging galaxies}
 
\author{Sandor Van Wassenhove$^{1}$, Marta Volonteri$^{1}$, Lucio Mayer$^{2}$, Massimo Dotti$^{3}$, Jillian Bellovary$^{1}$, Simone Callegari$^{2}$}
\affil{1 Department of Astronomy, University of Michigan, Ann Arbor, MI 48109, USA}
\affil{2 Institute of Theoretical Physics, University of Z\"urich, Winterthurerstrasse 190, CH-9057 Z\"urich, Switzerland}
\affil{3 Dipartimento di Fisica G. Occhialini, Universit\`a degli Studi di Milano Bicocca, Piazza della Scienza 3, 20126 Milano, Italy}

\begin{abstract}
Supermassive black holes (SMBHs) have been detected in the centers of most nearby massive galaxies. Galaxies today are the products of billions of years of galaxy mergers, but also billions of years of SMBH activity as active galactic nuclei (AGNs) that is connected to galaxy mergers. In this context, detection of AGN pairs should be relatively common. Observationally, however, dual AGN are scant, being just a few percent of all AGN. In this Letter we investigate the triggering of AGN activity in merging galaxies via a suite of high resolution hydrodynamical simulations. We follow the dynamics and accretion onto the SMBHs as they move from separations of tens of kiloparsecs to tens of parsecs. Our resolution, cooling and star formation implementation produce an inhomogeneous, multi-phase interstellar medium, allowing us to accurately trace star formation and accretion onto the SMBHs. We study the impact of gas content, morphology, and mass ratio, focusing on AGN activity and dynamics across a wide range of relevant conditions. We test when the two AGN are simultaneously detectable, for how long and at which separations. We find that strong dual AGN activity occurs during the late phases of the mergers, at small separations ($<$1-10 kpc) below the resolution limit of most surveys. Much of the SMBH accretion is not simultaneous, limiting the dual AGN fraction detectable through imaging and spectroscopy to a few percent, in agreement with observational samples. \end{abstract}

\keywords{galaxies: nuclei --- galaxies: interactions --- galaxies: active}

\section{Introduction}
The exceptional spatial resolution of the Hubble Space Telescope and of  Chandra led to the exciting discovery of the first three cases of dual active galactic nuclei (AGNs) in the center of the same galaxy, in LBQS~0103-2753, NGC~6240 and Arp~299 \citep{Junkkarinen2001,Komossa2003,Ballo2004}.  Recently, more observations were focused on detecting spatially resolved dual AGN via various techniques \citep{Gerke2007, Smith2010, Hudson2006, McGurk2011, Rodriguez2006, Bianchi2008, Barth2008, Comerford2009a,Liu2010,Piconcelli2010, Comerford2009b,Green2010,Liu2011,Fu2011,Koss2010,Koss2011,Fabbiano2011}. 

If most galaxies host a SMBH \citep{Ferrarese2005} and galaxy mergers trigger quasar activity \citep[e.g.,][]{DiMatteo2005} then one expects that dual AGNs should be common.   Observationally, however, AGN pairs are rare (at most a few percent, see references above). The prediction of the timescale on which SMBH pairs can be observed as double quasars \citep[and references therein]{Hennawi2006, Liu2011} is a key diagnostic of SMBH merger rates (of paramount importance for gravitational wave searches) and AGN triggering.

\cite{Foreman2009} discuss how the relationship between the lifetime of an active SMBH, $t_{\rm AGN}$, and the merging timescale, $ t_{\rm merg}$, plays a fundamental role in determining the observability of AGN pairs. If one assumes that most galaxies host SMBHs, that AGN/quasar activity is triggered by galaxy mergers, and that the lifetime of quasars equals the merger timescale, $t_{\rm AGN}\simeq t_{\rm merg}$, the probability of observing a dual quasar should be close to unity, if we do not consider additional factors, such as obscuration. If $t_{\rm AGN}\ll t_{\rm merg}$ \citep{DiMatteo2005, hopkins2005}, and/or if there is a delay in the triggering of the two quasars, then one might have ceased its activity before the other started.  \cite{Foreman2009} also notice that the distribution of physical separations for luminous quasar pairs in the Sloan Digital Sky Survey peaks below 30 kpc, the lower limit of the physical resolution that can be resolved in the survey. The paucity of optically selected quasar pairs on galactic scales \citep[$\sim 0.1$\% at $L \gta 10^{45}$ erg s$^{-1}$,][]{Foreman2009,Hennawi2006} points toward non-simultaneous activity at large separations.  

At lower levels of activity, \cite{Comerford2009a,Liu2010,Shen2011} find that about 2-5\% of optically-selected (via H$\beta$ and [OIII] lines) AGNs are in pairs, and 30\% off-set from the host center (hinting at inspiral).  On the other hand,  about 30\% of X-ray detected sub-mm galaxies at $z\approx2$ are in pairs \citep{Alexander2003}, suggesting that the occurrence of AGN pairs is possibly underestimated by optical selection. However, some observed AGN pairs are likely to be caused by gas kinematics rather than true dual AGN \citep{Shen2011, Fu2012}.

 \cite{Michiganders} and \cite{Yu2011} also discuss models that reconcile theoretical merger rates of SMBHs and galaxies with the small fraction of binary quasars and dual AGN respectively. The lifetime of AGN, the gas content of the host galaxies and the dynamics of the merger are the main factors that explain the paucity of observed AGN pairs.  Understanding the occurrence of AGN pairs therefore requires a thorough investigation and understanding of the detailed  physical  conditions describing the evolution of SMBHs during mergers.
 
In this Letter we investigate the theoretical expectations for detections of dual AGNs.  We use high resolution simulations to study the dynamics, statistics, and observability of AGN pairs to test if the scarce number of quasar/AGN pairs is consistent with limited observability of simultaneous activity of SMBHs involved in galaxy mergers.

\section{Simulations}

We investigate the dynamical and accretion history of SMBHs in merging galaxies via a suite of very high resolution ($<20$ pc) smoothed particle hydrodynamics (SPH) simulations. We present here two new simulations with mass ratios 1:2, one between two gas-rich spiral galaxies and the other between a gas-poor elliptical and a gas-rich spiral. We supplement our analysis with a 1:10 mass ratio simulation between gas-rich spirals from \cite{Callegari2011}. This suite focuses on galaxy mergers most relevant to the build up of a galaxy in the $\Lambda$CDM cosmology. As mergers between equal mass galaxies are rare, we probe the range of mergers that are more common but are expected to lead to efficient SMBH pairing. The peak of the cosmic merger rate occurs at higher redshift, so we begin our simulations at $z=3$. We include the elliptical-spiral merger to study the impact of gas fraction and morphology on SMBH pairing and accretion. This range of parameters allows us to study SMBH and host dynamics and co-evolution across a range of cosmologically relevant merger conditions. 

The methodology of how our spiral galaxies are initialized, as well as the details of the 1:10 merger, is described in \cite{Callegari2011}. We briefly summarize here the galaxy models that we use in the 1:2 mergers. The spiral galaxies contain a dark matter halo, a disk composed of stars and gas, and a stellar bulge. The halo is represented by a spherical \cite{NFW1996} profile with spin parameter $\lambda=0.04$. An exponential disk of stars and gas is included with total mass $0.04 M_{\rm vir}$ and gas fraction $f_g = 0.3$. We note that observations of more actively star forming galaxies at high redshift suggest even higher gas fractions \citep{Tacconi2010}. The stellar bulge is represented by a spherical \cite{Hernquist1990} model with mass $0.008 M_{\rm vir}$. The primary galaxy in the 1:2 and 1:10 spiral-spiral mergers and the secondary galaxy in the elliptical-spiral merger have $M_{\rm vir} = 2.24 \times 10^{11} M_{\odot}$. The elliptical galaxy consists of a dark matter halo and a stellar component, each represented by a \cite{Hernquist1990} profile. The halo has spin parameter $\lambda=0.04$ and scale length 22 kpc, chosen to resemble a \cite{NFW1996} profile following the method in \cite{springel2005b}, assuming concentration $c=3$. The stellar component has total mass $0.05 M_{\rm vir}$, no rotation, and scale length 0.5 kpc, chosen based on the scale length of early type galaxies from SDSS \citep{Shen03}.

In both 1:2 mergers represented here, dark matter particles have masses of $1.1\times 10^5 M_{\odot}$ and softening length 30 pc, gas particles have masses of $4.6\times 10^3 M_{\odot}$ and softening length 20 pc, and star particles have masses of $3.3\times 10^3 M_{\odot}$ and softening length 10 pc. This resolution allows us to consistently track the dynamics and evolution of the SMBHs down to the formation of a SMBH pair while minimizing their excursion from the centers of their galaxies. The high resolution also enables us to track the dynamics of gas in the galaxies, leading to accurate determinations of SMBH accretion rates, feedback, and star formation during different stages of the merger. Black hole masses are chosen to be consistent with the $M_{\rm BH}-M_{\rm bulge}$ relation \citep{MarconiHunt2003}. A single particle representing the SMBH is placed at the center of each galaxy. The initial SMBH masses are $4\times 10^7 M_{\odot}$ and $3\times 10^6 M_{\odot}$ in the elliptical-spiral merger and $3\times 10^6 M_{\odot}$ and $1.5\times 10^6 M_{\odot}$ in the 1:2 spiral-spiral merger for the primary and secondary SMBHs, respectively. For reference, the SMBH masses in the 1:10 merger from \cite{Callegari2011} are $6\times 10^5 M_{\odot}$ and $6\times 10^4 M_{\odot}$.

Each simulation begins with the galaxies at a separation equal to the sum of their virial radii. Orbital parameters were chosen based on results from cosmological simulations \citep{Benson05}. In the 1:2 elliptical-spiral merger, the galaxies have initial eccentricity 0.98 and pericentric distance 19\% of the virial radius of the primary galaxy. In the 1:2 spiral-spiral merger, the galaxies have initial eccentricity 1.02 and pericentric distance 26\% of the virial radius of the primary galaxy. All mergers are planar and prograde. We note that this orientation may produce stronger gas inflow than inclined mergers, likely maximizing SMBH accretion and, therefore, dual SMBH activity.

\begin{table*}
\begin{center}
\label{obsfraction}
\begin{tabular}{ccccccccc}
\hline
Simulation & Threshold & BH$_1$ & BH$_2$ & Dual & $d > 1$ kpc & $d > 10$ kpc & $\Delta v > 150$ km s$^{-1}$ \\
 & & & & AGN &(imaging) & (imaging) & (spectroscopy) \\
\hline
1:2 Spiral-Spiral     &  $L_{bol} > L_{\rm 42}$  &  $77.8$  &  $64.9$  &  $57.6$  &  $53.4$  &  $43.9$  &  $35.7$  \\
                      &  $L_{bol} > L_{\rm 43}$  &  $21.2$  &  $15.3$  &  $16.3$  &  $13.5$  &  $5.61$  &  $8.23$  \\
                      &  $L_{bol} > L_{\rm 44}$  &  $2.68$  &  $2.61$  &  $19.2$  &  $16.5$  &  $0.10$  &  $4.76$  \\
                      &  $f_{Edd} > 0.005$    &  $59.9$  &  $63.0$  &  $49.3$  &  $45.9$  &  $36.8$  &  $31.1$  \\
                      &  $f_{Edd} > 0.05$     &  $11.1$  &  $14.2$  &  $13.8$  &  $11.8$  &  $3.99$  &  $6.51$  \\
                      &  $f_{Edd} > 0.5$      &  $1.47$  &  $2.26$  &  $18.5$  &  $17.2$  &  $0.07$  &  $3.44$  \\
\hline
1:2 Elliptical-Spiral &  $L_{bol} > L_{\rm 42}$  &  $30.7$  &  $67.0$  &  $31.4$  &  $29.3$  &  $16.4$  &  $23.3$  \\
                      &  $L_{bol} > L_{\rm 43}$  &  $26.5$  &  $19.2$  &  $23.6$  &  $22.2$  &  $8.16$  &  $16.9$  \\
                      &  $L_{bol} > L_{\rm 44}$  &  $12.1$  &  $1.90$  &  $6.26$  &  $5.94$  &  $0.60$  &  $3.83$  \\
                      &  $f_{Edd} > 0.005$    &  $21.2$  &  $47.9$  &  $23.8$  &  $22.1$  &  $9.28$  &  $16.0$  \\
                      &  $f_{Edd} > 0.05$     &  $6.95$  &  $9.91$  &  $11.0$  &  $10.2$  &  $1.62$  &  $6.89$  \\
                      &  $f_{Edd} > 0.5$      &  $0.38$  &  $0.64$  &  $1.27$  &  $1.27$  &  $0.55$  &  $0.73$  \\
\hline
1:10 Spiral-Spiral    &  $L_{bol} > L_{\rm 42}$  &  $37.0$  &  $9.61$  &  $8.12$  &  $6.70$  &  $3.58$  &  $3.76$  \\
                      &  $L_{bol} > L_{\rm 43}$  &  $2.26$  &  $1.55$  &  $7.19$  &  $0.91$  &  $0$     &  $0.86$  \\
                      &  $f_{Edd} > 0.005$    &  $55.5$  &  $46.1$  &  $33.3$  &  $32.2$  &  $26.8$  &  $13.9$  \\
                      &  $f_{Edd} > 0.05$     &  $4.22$  &  $10.5$  &  $4.06$  &  $2.20$  &  $1.59$  &  $1.44$  \\
                      &  $f_{Edd} > 0.5$      &  $0.53$  &  $1.39$  &  $7.87$  &  $0.29$  &  $0$     &  $0.37$  \\
\hline
\end{tabular}
\end{center}
\caption{BH$_1$ and BH$_2$ represent the percentage of the total simulated time that the primary and secondary SMBHs are active, respectively, above the given activity threshold ($L_{\rm 42}\equiv 10^{42}$ erg s$^{-1}$; $L_{\rm 43}\equiv 10^{43}$ erg s$^{-1}$; $L_{\rm 44}\equiv 10^{44}$ erg s$^{-1}$; accreted mass at each timestep is converted into a bolometric luminosity assuming a radiative efficiency of 10\%). The total simulated time for each merger is: 1.35 Gyr for the 1:2 Spiral-Spiral simulation, 1.1 Gyr for the 1:2 Elliptical-Spiral simulation, and 2.56 Gyr for the 1:10 Spiral-Spiral simulation. The Dual AGN column gives the percentage of time that both SMBHs are active (over the total time that one or more SMBHs are active at the given threshold). The remaining columns give the dual activity percentage with additional constraints, based on imaging or spectroscopic detectability. $d$ represents the absolute separation between the SMBHs. $\Delta v$ is the absolute velocity difference between the SMBHs. }
\end{table*}

We performed our simulations using GASOLINE, a SPH $N$-body Tree code \citep{Wadsley04,Stadel01}. GASOLINE includes a physically motivated prescription for star formation and supernova feedback \citep{Stinson2006}, as well as a recipe for black hole physics \citep{Bellovary10}. We have chosen our parameters\footnote{The new simulations presented in this Letter use an updated version of GASOLINE. We refer the reader to \cite{Callegari2011} for a list of parameters used in the 1:10 simulation.} to match those in the literature for simulations of comparable resolution , which are able to produce realistic galaxies \citep[e.g.][]{Governato10}. Stars are able to form if the parent gas particle reaches a threshold density of 100 amu cm$^{-3}$ and is below a temperature of 6000 K. Supernovae release $E_{SN} = 10^{51}$ erg into the surrounding gas based on the blastwave formalism of \citet{Stinson2006}. In order to prevent an unphysical burst of supernovae as the simulation begins, we relax the galaxies in isolation, gradually increasing the star formation efficiency, $c^*$, from 0.005 to 0.015 over $10^8$ years. Black holes accrete gas through Bondi-Hoyle accretion. This accretion gives rise to thermal feedback, which we model as $\dot E = \epsilon_f \epsilon_r \dot M c^2$ \citep{springel2005b}, with radiative efficiency $\epsilon_r = 0.1$ and feedback efficiency $\epsilon_f = 0.001$ as in \citet{Bellovary10}. The feedback energy is imparted on the nearest gas particle to the black hole.

\section{Results}

\begin{figure}
\includegraphics[width= \columnwidth]{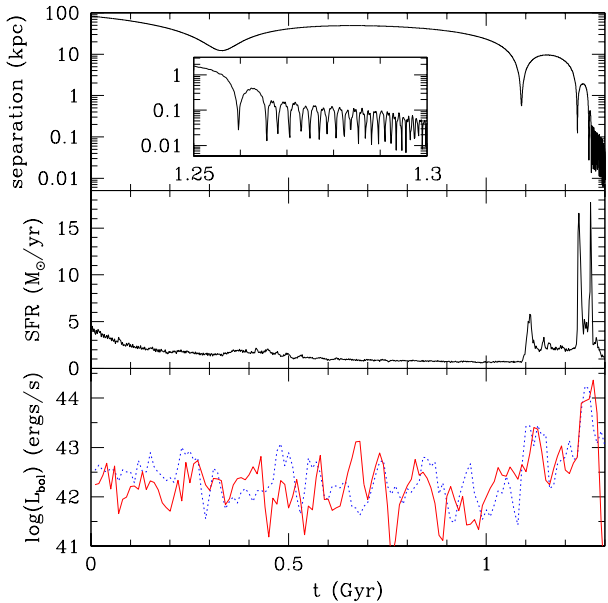}
\caption{Evolution of the 1:2 spiral-spiral merger. Upper panel: separation of the two black holes as a function of time. The inset shows a zoomed-in view of the separation at late times. Middle panel: star formation rate from both galaxies as a function of time. Lower panel: bolometric luminosities of the two black holes as a function of time. The blue, dotted line represents the primary SMBH. The red, solid line represents the secondary SMBH.}
\label{sepsfrL_SpSp}
\end{figure}

In Table 1, we present the results of our simulations. We determine the fraction of the total simulated time that each SMBH is active, given a threshold for observability. We also show the time spent as a dual AGN, when both SMBHs exceed the given threshold at the same time, expressed as a fraction of the total time that one or both SMBHs are active at that threshold. The dual AGN fraction is also given with additional constraints to mimic observational limitations. The separation thresholds represent possible spatial resolution cutoffs for separating the AGN with imaging. Spectroscopic duals require sufficiently large velocity offsets that two sets of emission lines are discernible. We note that these cutoffs are absolute, not projected, quantities in our analysis; they are meant to show the qualitative effects of observational limits.

{\bf 1:2 Spiral-Spiral merger.} We show in Fig.~\ref{sepsfrL_SpSp} the evolution of the SMBH separation, host galaxy star formation, and SMBH accretion. Star formation rates include stars formed in both the primary and secondary galaxies. Before the second pericenter passage, the SMBHs accrete relatively little, increasing in mass by 20\%-30\% over a Gyr of evolution. Accretion is not well correlated between the SMBHs, and bolometric luminosities generally remain at $<10^{43}$ erg s$^{-1}$. Following the second pericenter passage, however, tidal torques concentrate gas in the centers of the two galaxies, leading to efficient SMBH accretion accompanied by strong star formation, which may complicate detection of the AGN. Over the next 200 Myr, the SMBHs increase in mass by factors of 30\%-60\%. During this phase, when the SMBHs are separated by less than 10 kpc, the AGN bolometric luminosities reach $10^{44}$ erg~s$^{-1}$, and the activity is better correlated, triggering after each pericenter passage. These luminosities are sustained for $\simeq 35$ Myr by each SMBH separately, and for $\simeq 10$ Myr simultaneously. We end this simulation at $t=1.35$ Gyr, as the SMBHs have reached our resolution limit at separations of 10 pc. 

\begin{figure*}
\includegraphics[width=7in]{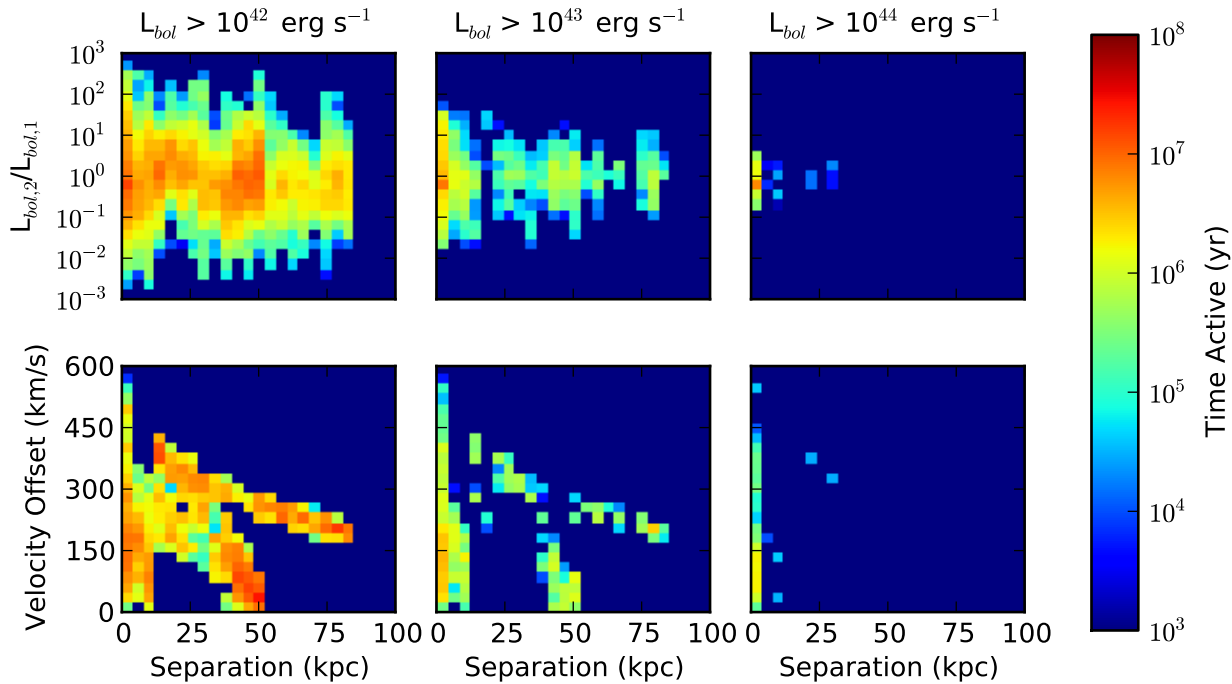}
\caption{Dual AGN observability timescale at a given SMBH separation, activity threshold, and luminosity ratio or velocity offset for the 1:2 spiral-spiral merger. In each case, the threshold is imposed upon both SMBHs.}
\label{observability_SpSp}
\end{figure*}

Fig.~\ref{observability_SpSp}\footnote{Additional figures for the other simulations and with log-scale separations are available at http://www.astro.lsa.umich.edu/\texttt{\char`\~}svanwas/dualAGN.html} shows the amount of time both SMBHs are active at a given separation and luminosity ratio for different thresholds. At low luminosity, the SMBHs are active for 60\%-70\% of the simulation and the observability of dual AGN activity traces the orbit of the host galaxies. There is significant activity at large separations as the galaxies spend most of their time at or near apocenter. However, a higher activity threshold selects for the phases of the merger where AGN triggering is strongest. The dual AGN activity at higher thresholds occurs mainly at small separations ($<$ 10 kpc), following the second and subsequent pericenter passages. We show also the distribution of velocity offsets between the SMBHs during dual AGN activity. The longest episode of dual activity above $10^{44}$ erg s$^{-1}$ occurs following the third pericenter passage, below separations of $\simeq 2$ kpc (see inset of Fig.~\ref{sepsfrL_SpSp}),  near apocenter. The velocity offset is small, therefore, and we find that $\simeq 75$ percent of the dual activity is at $\Delta v < 150$ km s$^{-1}$. In 1:1 mergers, \cite{DiMatteo2005} and \cite{hopkins2005} also find that the strongest SMBH activity occurs at small separations, although they do not distinguish between single and dual AGN activity. 

\begin{figure}
\includegraphics[width= \columnwidth]{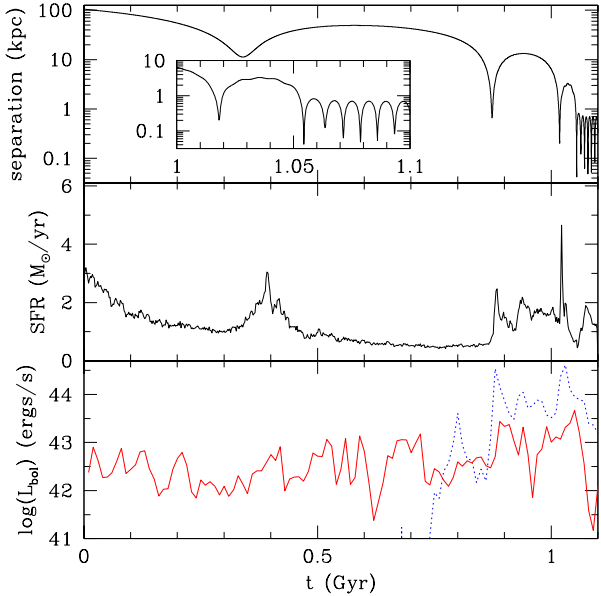}
\caption{Evolution of the 1:2 elliptical-spiral merger. Panels and symbols as in Fig.~1.}
\label{sepsfrL_ElSp}
\end{figure}

{\bf 1:2 Elliptical-Spiral merger.} In the 1:2 elliptical-spiral merger, the elliptical galaxy initially contains no gas. However, gas stripped from the companion galaxy during the first and second pericenter passages cools and settles at the center of the elliptical, allowing the primary SMBH to begin accreting at around $t=0.7$ Gyr (Fig.~\ref{sepsfrL_ElSp}). Since the primary SMBH is an order of magnitude more massive than the secondary, it produces a higher luminosity once gas is available. As in the spiral-spiral merger, the strongest SMBH accretion occurs following the second and subsequent pericentric passages. Above $10^{44}$ erg s$^{-1}$ dual AGN activity occurs for $\simeq 9$ Myr, mostly under 10 kpc separations.  We stop this simulation at $t = 1.1$ Gyr. At this time, the stars and gas surrounding the secondary SMBH have been tidally disrupted, preventing further AGN activity. The SMBHs remain at a separation of hundreds of pc. The dynamical friction timescale at this separation, ignoring the effects of gas, is of order a few hundred Myr \citep{Colpi99}, therefore the two SMBHs will eventually form a binary. 

Comparing the 1:2 mergers at the same Eddington fraction threshold yields insight into the AGN triggering in the two simulations. At all Eddington fractions, the SMBHs spend more time active in the spiral-spiral merger. Luminosity thresholds favor detectability in the elliptical-spiral merger because of the larger SMBH masses. We note that we have not included 
gas in the elliptical galaxy in the form of a hot halo. Ram pressure from the hot halo could strip gas from the spiral galaxy during the merger, preventing the secondary SMBH from accreting. Additionally, the hot halo could prevent stripped gas from cooling and reaching the primary SMBH. We expect the AGN timescales from our elliptical-spiral merger, therefore, to be upper limits.

{\bf 1:10 Spiral-Spiral merger.} We perform the same analysis on the 1:10 spiral-spiral merger with f$_g$ = 0.3 from Callegari et al. (2011, cf. analogs of Fig.~1 for this simulation). This merger is longer than the 1:2 mergers and includes many more pericenter passages and stronger stripping of the secondary galaxy. The primary galaxy is relatively undisturbed, leaving the primary SMBH to grow quiescently. The secondary SMBH, on the other hand, accretes at a higher Eddington fraction until ram pressure strips all gas from the secondary galaxy. When the secondary galaxy enters the disk of the primary and its orbit circularizes, the secondary SMBH begins accreting again, leading to a phase of dual AGN activity at sub-kpc separations. Table 1 shows that the majority of dual AGN activity at high activity thresholds occurs at separations under 1 kpc. 
Overall, the 1:10 spiral-spiral merger yields less dual AGN activity than the 1:2 mergers, even though it is a factor of two longer in duration. Stronger tidal effects and ram pressure on the secondary galaxy mean that the secondary SMBH is unable to accrete efficiently for a significant amount of time above kpc-scale separations.  The AGN reach bolometric luminosities of $\simeq 10^{43}$ erg s$^{-1}$ for $\simeq 60$ Myr and $\simeq 40$ Myr for the primary and secondary SMBHs respectively, and for $\simeq 7$ Myr simultaneously.

\section{Discussion and Conclusions}

We perform numerical simulations of galaxy mergers, focusing on the separations and timescales for dual AGN activity. In all of the simulations presented here, much of the AGN activity is non-simultaneous at high activity thresholds. We find that the dual activity is generally a small fraction of the total AGN activity, with each SMBH accreting longer alone than in a pair. At our lowest thresholds, the long dual AGN timescales simply reflect the almost non-stop low level accretion onto the SMBHs. At higher thresholds we are instead probing accretion that is triggered by the dynamics of the merger, yielding better correlated accretion between the SMBHs. However, significant non-simultaneous accretion remains.

Our results can be summarized as follows:
\begin{itemize}
\item At high luminosity thresholds, almost all dual AGN activity occurs at separations $< 10$ kpc, where AGN triggering is strongest. Lower thresholds favor instead large separations, where the galaxies spend more time during the merger.
\item Much of the AGN activity during the mergers is non-simultaneous. SMBHs are active alone rather than as a pair for $\sim 90\%$ of the time at $L_{bol} > 10^{44}$ erg s$^{-1}$  \citep[corresponding to an optical luminosity of order $10^{43}$ erg s$^{-1}$,][]{Marconi2004} , separations $>1-10$ kpc and $\Delta v > 150$ km s$^{-1}$. These SMBHs will appear as either `normal' single AGN (at low spectral or spatial resolution) or as offset AGN. This is in agreement with \cite{Comerford2009a}, who find that it is more common to find offset AGN rather than dual AGN. 
\item  From Table 1, the expected dual AGN fraction with $L_{bol} > 10^{44}$ erg s$^{-1}$ is ~4-5\% in our 1:2 mergers if we consider sufficiently luminous dual AGN that can be spectroscopically identified ($\Delta v > 150$ km s$^{-1}$). While our results are not directly comparable to observations of dual AGN, which typically occur at $z<1$, the dual AGN fraction is in broad agreement with that found by \cite{Shen2011} in SDSS AGN. Less than 1\% of luminous dual systems would be identified via imaging in surveys with spatial resolution less than 10 kpc \citep{Foreman2009}.
\item The mass ratio and morphologies of merging galaxies determine when their SMBHs will be fed and starved. Tidal stripping unbinds the stars and gas surrounding the secondary SMBH in our 1:2 elliptical-spiral merger, preventing strong dual activity at sub-kpc separations. In the 1:10 spiral-spiral merger, the secondary galaxy is stripped of its gas during the fourth pericenter passage. 
\end{itemize}

We note that these estimate of dual AGN activity are upper limits as we have not included dust and obscuration in our models, nor dilution by star formation. This may hinder detection of the AGN \citep{Sanders88,hopkins2005}, especially in optical surveys, but perhaps not in hard X-rays \citep{Koss2011}. As we do not follow the SMBHs past our resolution limit, we do not include binary AGN or post-coalescence activity in our analysis. Additionally, there is likely to be considerable variability in the SMBH acccretion rates on smaller timescales and spatial scales than we resolve \citep{HQ2010, Levine2010}. This unresolved variability could decrease the overall dual fraction as more accretion becomes non-simultaneous. We will address detectability in different bands in a follow-up paper, taking into account the luminosity of the host and dust obscuration.

\begin{acknowledgments}

The authors are grateful to the anonymous referee for comments that improved the clarity of the Letter. Simulations were run using computer resources and technical support from NAS. MV acknowledges support from SAO Award TM1-12007X, NASA award ATP NNX10AC84G and NSF award AST 1107675.

\end{acknowledgments}


\end{document}